\documentclass{svjour3}
\usepackage{graphicx}
\usepackage{natbib}
\bibpunct{(}{)}{;}{a}{}{,}
\usepackage{amsmath,amssymb}      

\journalname{SSRv}

\usepackage{epsfig}

\newcommand{\be}{\begin{equation}}
\newcommand{\ee}{\end{equation}}
\newcommand{\beq}{\begin{eqnarray}}
\newcommand{\eeq}{\end{eqnarray}}

\def\alf{Alfv\'en~}

\def \xmm {{\sl XMM-Newton}}

\def \degmark{^\circ}

\newcommand{\ion}[2]{#1\,{\sc{#2}}}

\begin{document}

\title{Cosmological shock waves}

\author{A.M. Bykov  \and
         K. Dolag \and
        F. Durret}

\institute{A.M. Bykov \at A.F. Ioffe Institute of Physics and
Technology, St. Petersburg,
           194021, Russia\\
 \email{byk@astro.ioffe.ru} \and
K. Dolag \at Max-Planck-Institut f\"ur Astrophysik, P.O. Box 1317,
D-85741 Garching, Germany \\
\email{kdolag@mpa-garching.mpg.de} \and
           F. Durret \at  Institut d'Astrophysique de Paris, CNRS, Universit\'e
           Pierre et Marie Curie, 98bis Bd Arago, F-75014 Paris,
    France \\
\email{durret@iap.fr}
             }

\date{Received: 17 September 2007; Accepted: 15 October 2007}

\maketitle

\begin{abstract}
Large-scale structure formation, accretion and merging processes,
AGN activity produce cosmological gas shocks. The shocks convert a
fraction of the energy of gravitationally accelerated flows to
internal energy of the gas. Being the main gas-heating agent,
cosmological shocks could amplify magnetic fields and accelerate
energetic particles via the multi-fluid plasma relaxation processes.
We first discuss the basic properties of standard single-fluid
shocks. Cosmological plasma shocks are expected to be collisionless.
We then review the plasma processes responsible for the microscopic
structure of collisionless shocks.  A tiny fraction of the particles
crossing the shock is injected into the non-thermal energetic
component that could get a substantial part of the ram pressure
power dissipated at the shock. The energetic particles penetrate
deep into the shock upstream producing an extended shock precursor.
Scaling relations for postshock ion temperature and entropy as
functions of shock velocity in strong collisionless multi-fluid
shocks are discussed. We show that the multi-fluid nature of
collisionless shocks results in excessive gas compression, energetic
particle acceleration, precursor gas heating, magnetic field
amplification and non-thermal emission. Multi-fluid shocks provide a
reduced gas entropy production and could also modify the observable
thermodynamic scaling relations for clusters of galaxies.

\end{abstract}

\section{Introduction}
\label{Introduction}

The observed large scale structure of the Universe is thought to be
due to the gravitational growth of density fluctuations in the
post-inflation era. In this model, the evolving cosmic web is
governed by non-linear gravitational growth of the initially weak
density fluctuations in the dark energy dominated cosmology. The web
is traced by a tiny fraction of luminous baryonic matter.
Cosmological shock waves are an essential and often the only way to
power the luminous matter by converting a fraction of gravitational power
to thermal and non-thermal emissions of baryonic/leptonic matter.

At high redshifts ($z> 1100$) the pre-galactic medium was hot,
relatively dense, ionised, with a substantial pressure of radiation.
The cosmic microwave background (CMB) observations constrain the
amplitudes of density inhomogeneities to be very small at the last
scattering redshift $z \sim$ 1000. Strong non-linear shocks are
therefore unlikely at that stage. The universe expands, the matter
cools, and eventually recombines, being mostly in neutral phase
during the "dark ages" of the universe. At some redshift, $6 < z <
 14$, hydrogen in the universe is reionised, likely due to UV
radiation from the first luminous objects, leaving the intergalactic
medium (IGM) highly reionised (see e.g. \citet{Fan_ea06} for a recent
review). The reionisation  indicates the formation of the first
luminous objects at the end of the "dark ages", either star-forming
galaxies or Active Galactic Nuclei (AGN). The compact luminous
objects with an enormous energy release would have launched strong
(in some cases, relativistic) shock waves in the local vicinity of
the energetic sources. At the same evolution stage, formation of
strong density inhomogeneities in the cosmic structure occurs. Since
then the non-linear dynamical flows in the vicinity of density
inhomogeneities  would have created large scale cosmic structure
shocks of modest strength, thus heating the baryonic matter and
simultaneously producing highly non-equilibrium energetic particle
distributions, magnetic fields and electromagnetic emission.

Most of the diffuse X-ray emitting matter was likely heated by
cosmological shocks of different scales. Accretion and merging
processes produce large-scale gas shocks. Simulations of structure
formation in the Universe predict that in the present epoch about
40~\% of the normal baryonic matter is in the Warm-Hot Intergalactic
Medium (WHIM) at overdensities $\delta \sim 5-10$ (e.g.
\citealt{CenO99,Dave_ea01}). The WHIM is likely shock-heated
to temperatures of 10$^5 -- 10^7$~K during the continuous
non-linear structure evolution and star-formation processes.

The statistics of cosmological shocks in the large-scale structure
of the Universe were simulated in the context of the $\Lambda$CDM-cosmology using PM / Eulerian adiabatic hydrodynamic codes (e.g.
\citealt{Miniati_ea00,Ryu_ea03,Kang_ea07}) and
more recently with a smoothed particle hydrodynamic code by
\citet{Pfrommer_ea06}. They identified two main populations of
cosmological shocks: $(i)$ high Mach number "external" shocks due to
accretion of cold gas on gravitationally attracting nodes, and
$(ii)$ moderate Mach number ($2 \leq {\cal M}_{\rm s} \leq 4$)
"internal" shocks. The shocks are due to supersonic flows induced by
relaxing dark matter substructures in relatively hot, already
shocked, gas. The internal shocks were found by \citet{Ryu_ea03} to
be most important in energy dissipation providing intercluster
medium (ICM) heating, and they were suggested by
\citet{Bykov_ea00cl} to be the likely sources of non-thermal
emission in clusters of galaxies.

Hydrodynamical codes deal with N-body CDM and single-fluid
gas dynamics. However, if a strong accretion shock is multi-fluid,
providing reduced post-shock ion temperature and entropy, then the
internal shocks could have systematically higher Mach numbers.

Space plasma shocks are expected to be {\sl collisionless}.
Cosmological shocks, being the main gas-heating agent, generate
turbulent magnetic fields and accelerate energetic particles via
collisionless multi-fluid plasma relaxation processes thus producing
non-thermal components. The presence of these non-thermal components
may affect the global dynamics of clusters of galaxies
\citet{Ostriker_ea05} and the $\sigma_{\rm v}$-$T$, $M$-$T$, $L_{\rm
X}$-$T$ scaling relations \citet{Bykov05}. Detailed discussion of
the cosmological simulations of the scaling relations with account
of only thermal components can be found in \citealt{13_borgani2008}
- Chapter 13, this volume.

 In Sect.~\ref{colls} we discuss the basic features of the standard collisional shocks.
The main part of the review is devoted to physical properties of
cosmological shocks with an accent on collisionless shocks and
associated non-thermal components. In Sect.~\ref{CRS} we discuss
the most important features of multi-fluid collisionless shocks in
the cosmological context including the effects of reduced entropy
production, energetic particle acceleration and magnetic field
amplification in the shocks.

\section{Single-fluid MHD-shocks}
\label{colls}

Shock waves are usually considered as a sharp transition between a
macroscopic supersonic (and super-Alfv\'enic) upstream flow (state 1)
and slowed down to a subsonic velocity downstream flow (state 2),
providing a mass flow $j_n$ through the shock surface. It is assumed
that a gas particle (or an elementary macroscopic fluid cell) is at
any instant of time in the {\sl local thermodynamic equilibrium
state} corresponding to the instantaneous values of the macroscopic
parameters. The Maxwellian distribution of all species is ensured
after a few molecular (or Coulomb) collisions have occurred. The
macroscopic parameters characterising the state of the gas, such as
density, specific internal energy, or temperature, change slowly in
comparison with the rates of the relaxation processes leading to
thermodynamic equilibrium. We consider here a single-fluid plasma
model assuming complete electron-ion relaxation. Under these
conditions, in a frame moving with the shock front, with the matter
flux across the shock surface $j_n \neq$0, the conservation laws for
mass (in non-relativistic flows), momentum and energy can be written
as follows:

\begin{equation}
j_n \;[\frac{\mathbf{B_t}}{\rho}] = B_n\,[\mathbf{u_t}]\, ,
\label{eq:cont1}
\end{equation}

\begin{equation}
j_n [\mathbf{u_t}] = \frac{B_n}{4\pi j_n}\,[\mathbf{B_t}]\, ,
\label{eq:cont2}
\end{equation}

\begin{equation}
[\;\frac{j_n^2}{\rho} + P + \frac{B_t^2}{8\pi}\;] =0\, ,
\label{eq:cont3}
\end{equation}

\begin{equation}
 [\;w + \frac{j_n^2}{2\rho^2}\, + \frac{u_t^2}{2}\, +
\frac{B_t^2}{4\pi \rho}\, -\frac{B_n}{4\pi
j_n}\,\mathbf{B_t}\mathbf{u_t}\;] =  0\, .
 \label{eq:cont4}
\end{equation}

\begin{figure}    
\begin{center}
\hbox{
\psfig{file=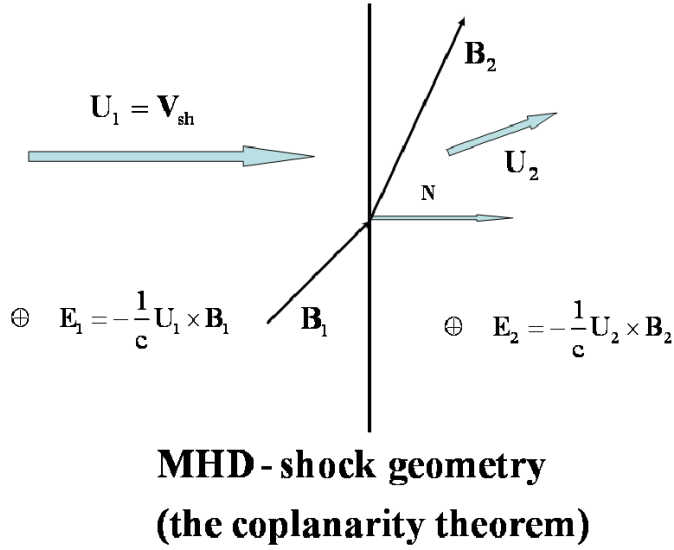,width=11.5cm,clip=}}
\caption{A sketch illustrating the coplanarity theorem  for a plane
ideal MHD-shock. The upstream and downstream bulk velocities
$\mathbf{U_1}$ and $\mathbf{U_2}$, magnetic fields $\mathbf{B_1}$
and $\mathbf{B_2}$ and the shock normal $\mathbf{N}$ all lie in the
same plane. The shock is at rest in the reference frame where also
$u_{t1}$ = 0. The shock is of infinitesimal width in the sketch.
Simulated structure of the transition region of a collisionless
shock is shown  in Fig.~\ref{hyb_B} and Fig.~\ref{hyb_phase}  where
its finite width is apparent.} \label{geometry}
\end{center}
\end{figure}

Here $\mathbf{U}=(u_n,\mathbf{u_t})$ is the bulk velocity, $w=
\varepsilon + P/\rho$ is the gas enthalpy, $\varepsilon\, , P, \rho
$ are the internal gas energy, pressure and density respectively.
The subscripts $n$ and $t$ are used for the normal and transverse
components respectively. We used the standard notations $[A] = A_2 -
A_1$ for the jump of a function $A$ between the downstream and
upstream regions. In the MHD case the relations equivalent to
Eq.~\ref{eq:cont1}$-$\ref{eq:cont4} were obtained by
\citet{deHoffmannT50}. The equations are valid in the case of a
magnetic field frozen-into moving plasma, where $\mathbf{E} =
-\frac{\mathbf{U}}{c}\times \mathbf{B}$.
 A specific feature of MHD shock waves is the
so-called coplanarity theorem (e.g. \citealt{LandauL84}) saying that
the upstream and downstream magnetic fields $\mathbf{B_1}$ and
$\mathbf{B_2}$ and the shock normal all lie in the same plane as it
is illustrated in Fig.~\ref{geometry}. It is important to note that
if $B_n \neq 0$ there is an especial reference frame where local
velocity $\mathbf{U}$ and magnetic field $\mathbf{B}$ are parallel
both in the upstream and downstream, providing $\mathbf{E}=0$.

From Eq.~\ref{eq:cont1}$-$\ref{eq:cont4} one may obtain a
generalised Rankine-Hugoniot (RH) adiabat

\begin{equation}
\varepsilon_2 - \varepsilon_1 + \frac{1}{2}(\frac{1}{\rho_2}
-\frac{1}{\rho_1})\{(P_2 + P_1) + \frac{1}{8 \pi}(B_{t2} -
B_{t1})^2)\} = 0.
  \label{eq:hugon}
\end{equation}

The RH adiabat connects the macroscopic parameters downstream of the
flow once the upstream state is known. In a parallel shock
($\mathbf{B}_t$=0)
\be r = \frac{\rho_{2}}{\rho_{1}} = \frac{(\gamma_{\rm g} +1) {\cal
M}_{\rm s}^2} {(\gamma_{\rm g} -1) {\cal M}_{\rm s}^2 + 2},
\label{eq:rhr} \ee

\begin{equation} \frac{T_2}{T_1} = \frac{[2 \gamma_{\rm g} {\cal
M}_{\rm s}^2 - (\gamma_{\rm g} -1)] [(\gamma_{\rm g} -1) {\cal
M}_{\rm s}^2 + 2]} {(\gamma_{\rm g} +1)^2 {\cal M}_{\rm s}^2},
\label{eq:rht} \end{equation} 
where $\gamma_{\rm g}$ is the gas
adiabatic exponent. We restrict ourselves here to a fast mode
shock where $c_{\rm s1} < u_{1}$, and $v_{\rm a2} < u_{2} < c_{\rm
s2}$, for $v_{\rm a1} < c_{\rm s1}$. The phase velocity $v_{\rm a2}$
 is the Alfv\'en velocity in the downstream, $c_{\rm s1}$, $c_{\rm
s2}$ are the sound speeds in the upstream and downstream
respectively. We define here the shock Mach numbers as ${\cal
M}_{\rm s} = v_{\rm sh}/c_{\rm s1}$ and ${\cal M}_{\rm a} = v_{\rm
sh}/v_{\rm a1}$.

In the case of a perpendicular shock ($B_n =0$) the compression
ratio is
\be r = \frac{\rho_{2}}{\rho_{1}} = \frac{2(\gamma_{\rm g} +1)} {
\psi + (\psi^2 + 4(\gamma_{\rm g} +1)\,(2- \gamma_{\rm g})\,{\cal
M}_{\rm a}^{-2})^{1/2}}, \label{eq:rhrp} \ee

$$
\psi = (\gamma_{\rm g} -1)\,+\,(2{\cal M}_{\rm s}^{-2} + \gamma_{\rm
g}\,{\cal M}_{\rm a}^{-2}).
$$
In a single fluid strong shock with ${\cal M}_{\rm s} \gg 1$ and
${\cal M}_{\rm a} \gg 1$ one gets
\begin{equation}
 T_{2} \approx 2\cdot \frac{(\gamma_{\rm g}
-1)}{(\gamma_{\rm g} +1)^2}~\mu v_{\rm sh}^2 = 1.38 \times
10^7~v^2_{\mathrm s8}~({\mathrm K}), \label{eq:rht1}
\end{equation}
 for any magnetic field
inclination (e.g. \citealt{DraineM93}).  The mass per particle $\mu$
was assumed to be $[1.4/2.3]  m_{\mathrm H}$ and $v_{\rm s8}$ is the
shock velocity in $10^8$~cm\,s$^{-1}$.

The RH adiabat does not depend on the exact nature of the
dissipation mechanisms that provide the transition between the
states 1 and 2. It assumes a single-fluid motion in regular
electromagnetic fields. However, the dissipative effects control the
thickness of the shock transition layer. In the case of a weak shock
of Mach number ${\cal M}_{\rm s} - 1 \ll 1$ the thickness is large
enough, allowing a macroscopic hydrodynamical description of the
fluid inside the shock transition layer (e.g. \citealt{LandauL84}).
The gas shock width $\Delta$ in collisional hydrodynamics without
magnetic fields is given by
\begin{equation}
\Delta = \frac{8 a V^2}{(P_2-P_1)\,(\partial^2{V}/\partial{P^2})_s}.
 \label{eq:width1}
\end{equation}
Following \citet{LandauL59} the gas shock width  in
Eq.~\ref{eq:width1} can be expressed through the viscosities $\eta$
and $\zeta$, and thermal conductivity $\kappa$, since
$$
a =\frac{1}{2\rho v_{\rm s}^3}[(\frac{4}{3}\eta + \zeta) +
\kappa(\frac{1}{c_{\rm v}} - \frac{1}{c_{\rm p}})]
$$
Here ${c_{\rm v}}$ and ${c_{\rm p}}$ are specific heats at constant
volume and at constant pressure respectively. Extrapolating
Eq.~\ref{eq:width1} to a shock of finite strength where $P_2-P_1
\sim P_2$, one may show that the gas shock width $\Delta$  is of the
order of the mean free path $\lambda$.

It is instructive to note that the entropy is non-monotonic inside
the finite width of a weak gas shock $({\cal M}_{\rm s} - 1) \ll 1$
and the total RH jump of the entropy $\Delta s$ across the shock is
of the third order in $({\cal M}_{\rm s} - 1)$:
\begin{equation}
\Delta s =  \frac{1}{12 T_1}
\Bigl( \frac{\partial^2{V}}{\partial{P^2}} \Bigr)_s\,  (P_2 - P_1)^3\,
\propto ({\cal M}_{\rm s} - 1)^3,
 \label{eq:entr}
\end{equation}
while the density, temperature and pressure jumps are $\propto
({\cal M}_{\rm s} - 1)$ \citep{LandauL59}.

In plasma shocks the shock structure is more complex because of a
relatively slow electron-ion temperature relaxation. Such a shock
consists of an ion viscous jump and an electron-ion thermal relaxation
zone. In the case of plasma shocks the structure of the ion viscous
jump is similar to the single fluid shock width structure discussed
above and can be studied accounting for the entropy of an isothermal
electron fluid. The shock ion viscous jump has a width of the order
of the ion mean free path. The scattering length (the mean free path
to $\pi$/2 deflection) $\lambda_{\mathrm p}$ of a proton of velocity $v_7$
(measured in 100~km\,s$^{-1}$) due to binary Coulomb collisions with
plasma protons of density $n$ (measured in cm$^{-3}$) can be estimated
as $\lambda_{\mathrm p} \approx 7 \times 10^{14}~ v_{7}^4\,  n^{-1}$ cm
\citep{Spitzer62}. After the reionisation ($z < 6$) the Coulomb mean
free path in the WHIM of overdensity $\delta$ is $\lambda_{\mathrm p}
\approx 3.5 \times 10^{21}~  v_{7}^4\,  \delta^{-1}\, 
(1+z)^{-3}  (\Omega_{\mathrm b}h^2/0.02)^{-1}$~cm. Here and below
$\Omega_{\mathrm b}$ is the baryon density parameter.
 The mean free path due to Coulomb collisions is typically
some orders of magnitude smaller than that for the charge-exchange
collisions in the WHIM after reionisation. The ion-electron thermal
relaxation occurs on scales about $\lambda_{\mathrm e}\times \sqrt{m_{\rm
p}/Z m_{\rm e}}$. Since $\lambda_{\mathrm e} \sim  \lambda_{\mathrm p}$, the width
of the relaxation zone is substantially  larger than the scale size
of the ion viscous jump. The application of the single fluid shock
model Eq.~\ref{eq:cont1}$-$\ref{eq:cont4} to electron-ion
plasmas assumes full ion-electron temperature relaxation over  the
shock width. For a discussion of the relaxation processes see e.g.
\citealt{bykov2008} - Chapter 8,  this volume, and references
therein.

In a rarefied hot cosmic plasma the Coulomb collisions are not
sufficient to provide the viscous dissipation of the incoming flow,
and collective effects due to the plasma flow instabilities play a
major role, providing the {\sl collisionless} shocks, as it is
directly observed in the heliosphere. The observed structure of
supernova remnants (e.g. \citealt{WeisskopfH06}) is consistent with
that expected if their forward shocks are collisionless. Moreover,
the non-thermal synchrotron emission seen in radio and X-rays is
rather a strong argument for high energy particle acceleration by
the shock that definitely favours its collisionless nature. That
allows us to suggest that cosmological shocks in a rarefied highly
ionised plasma (after the reionisation epoch) are likely to be
collisionless. There are yet very few observational studies of
cosmological shocks (e.g. \citealt{MarkevitchV07}). We review some
basic principles of collisionless shock physics in the next section.

\section{Collisionless shocks}
\label{colless}

Since the discovery of the solar wind in the early 1960's it has
been realised that the rapid rise time of magnetic storms observed
in the Earth suggested  very thin {\sl collisionless} shocks created
by solar flares (see for a discussion \citealt{Sagdeev66,Kennel_ea85}). The thickness of a viscous jump in a strong
{\sl collisional} shock is of the order of a mean free path (see
e.g. \citealt{ZeldovichR67}). The Coulomb collision mean free path in
the tenuous solar wind plasma is comparable to the Sun-Earth
distance, and thus the magnetic storm rising time due to standard
collisional shocks would exceed the observed time by orders of
magnitude.

 There are very specific features of collisionless
plasma shocks \citep{Sagdeev66}. Shocks in dense enough plasma
with frequent Coulomb collisions evolve very fast to Maxwellian
particle distributions with very few particles at high energies. On
the contrary, in collisionless plasma shocks, a small minority of
particles could gain a disproportionate share of the energy and
become non-Maxwellian. Collisionless shocks enable acceleration of a
small fraction of the particles to very high energies. Moreover, the
accelerated particles could carry away a substantial amount of the
kinetic energy of the plasma flow dissipated at the shock. The
energetic particles can penetrate far into the shock upstream gas,
to create an extended shock precursor. The cold gas in the shock
upstream is decelerated and pre-heated by the energetic particle and
MHD-wave pressure on a scale larger than a mean free path of an
energetic particle. This occurs not only at the bow shock of the
Earth at moderately low energies, but also in astrophysical shocks
at highly relativistic energies (e.g. \citealt{Russell05}).

A direct study of collisionless shock waves in a laboratory is an
extremely difficult task. Most of the experimental data on
collisionless shock physics are coming from space experiments. There
are direct observational data on the shock wave structure in the
interplanetary medium with clear evidence for ion and electron
acceleration by the shocks (e.g. \citealt{TsurutaniL85,Russell05}).

Computer simulations of the full structure of collisionless shock
waves describe the kinetics of multi-species particle flows and
magneto-hydrodynamic (MHD) waves in the strongly-coupled system. The
problem is multi-scale. It requires a simultaneous treatment of both
"microscopic" structure of the subshock at the thermal ion gyroradii
scale where the injection process is thought to occur, and an
extended "macroscopic" shock precursor due to energetic particles.
The precursor scale is typically more than $10^9$ times the
microscopic scale of the subshock transition region.

Energetic particles could be an essential component in the WHIM and
clusters of galaxies. Nonthermal particle acceleration at shocks is
expected to be an efficient process at different evolutional stages
of clusters.  Being the governing process of the supernova remnant
collisionless shock formation, nonlinear wave-particle interactions
are responsible for both shock heating and compression of the
thermal gas, as well as for creation of an energetic particle
population.


\subsection{Micro processes in collisionless shocks}

In the strong enough collisionless shocks (typically with a Mach
number above a few) resistivity cannot provide energy dissipation
fast enough to create a standard shock transition (e.g.
\citealt{Kennel_ea85}) on a microscopic scale. Ion instabilities are
important in such shocks that are called {\sl supercritical}.

At the microscopic scale the front of a supercritical shock wave
is a transition region occupied by magnetic field fluctuations of
an amplitude $\delta B/B \sim 1$ and characteristic frequencies of
about the ion gyro-frequency. Generation of the fluctuations is
due to instabilities in the interpenetrating multi-flow ion
movements. The width of the transition region of a quasi-parallel
shock wave reaches a few hundred ion inertial lengths defined as
$l_i = c/\omega_{\rm pi} \approx 2.3 \times 10^7 n^{-0.5}$~cm.
Here $\omega_{\rm pi}$ is the ion plasma frequency and $n$ is the
ionised ambient gas number density measured in cm$^{-3}$. The ion
inertial length in the WHIM can be
estimated as $l_i \approx 5.1 \times 10^{10}\,  \delta^{-1/2}
\, (1+z)^{-3/2}  (\Omega_{\mathrm b}h^2/0.02)^{-1/2}$~cm, providing
the width of the collisionless shock transition region is smaller
by many orders of magnitude than the Coulomb mean free path (that
is in the kiloparsec range).

The transition region of a quasi-perpendicular shock is somewhat
narrower. The wave generation effects at the microscopic scale have
been studied in some detail with hybrid code simulations (e.g.
\citealt{Quest88}).
The large-amplitude magnetic field fluctuations in the shock
transition region were directly measured in the interplanetary
medium (see e.g. \citealt{Kan_ea91}).

There are a few ways to simulate numerically the kinetics of the
collisionless plasma phenomena. The most comprehensive study of the
collisionless shock structure can be performed with the
particle-in-cell (PIC) method where all the plasma components are
considered as discrete particles in self-consistent fields. The PIC
method allows one to resolve electron scale lengths and frequencies,
but on the other hand it requires considerable computer resource. A
serious constraint on PIC and other plasma particle simulations of
collisionless shocks is that they must be done fully in three
spatial dimensions (3D). 
\citet{Jones_ea98} have proved that PIC simulations with one or more
ignorable dimensions artificially confine particles to field lines
and particularly eliminate cross-field diffusion. The effect is
especially important for simulations of a creation of a superthermal
particle population. All three box dimensions must be involved in
these simulations. Exact modelling of electron kinetics in
collisionless shocks require PIC simulations (e.g.
\citealt{HoshinoS02,Schmitz_ea02}). On the other hand, the
bulk of the energy of a collisionless shock is carried by the ions
and velocity relaxation processes are typically longer than the ion
gyro-periods. Thus, though the basic shock physics evolve on ion
spatial and temporal scales, the electron kinetic description
requires fine resolution at electron scales.

A fairly good description of low-frequency processes of the ion
dynamics in the shock transition layer can be achieved with hybrid
codes (e.g. \citet{WinskeO96} and references therein). Hybrid code
modelling, which interprets protons as particles and electrons as an
inertialess liquid, has made it possible to describe some important
features of the (sub)shock waves at the microscopic scale of some
hundred times the ion inertial length (e.g. \citealt{Quest88,Lembege_ea04,Burgess_ea05}).

A typical initialisation of a shock in the hybrid code simulations
is to inject a relatively cold ion beam (say at the right-hand
boundary) and to put a particle reflecting wall at the left-hand
boundary of a simulation box. In that case the shock is moving, and
the available simulation time is limited, given the finite size of
the box. The limited simulation time and the particle statistics per
cell are challenging the direct modelling of the origin and evolution
of the energetic non-thermal  particle population in a shock. To
increase the statistics the macro-particle splitting method is used
(see e.g. \citealt{Quest92,GiacaloneE00}).

\begin{figure}    
\begin{center}
\hbox{
\psfig{file=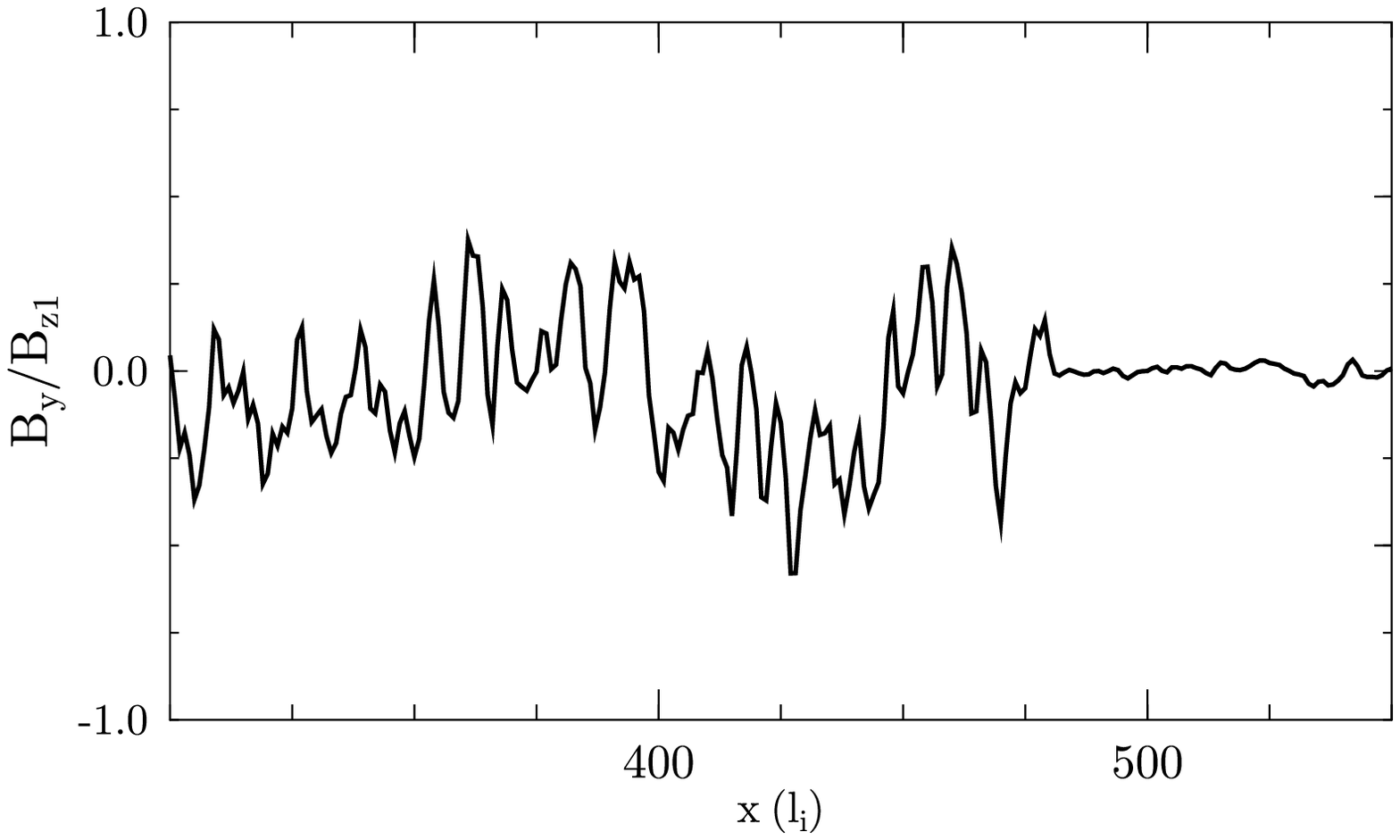,width=0.49\textwidth,clip=}
\psfig{file=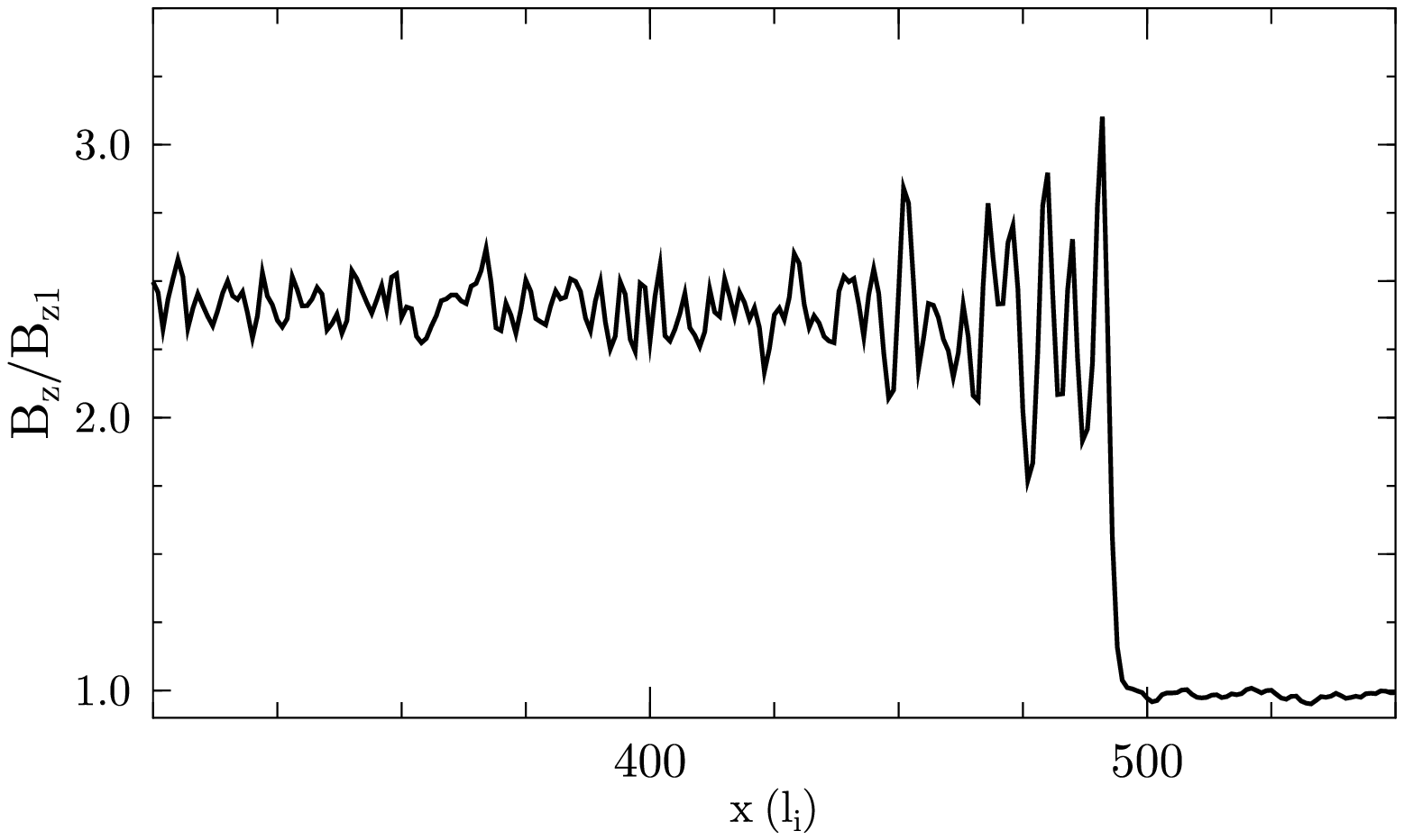,width=0.49\textwidth,clip=}}
\caption{Hybrid simulated magnetic fields of a quasi-perpendicular
shock (80$\degmark$ inclination). The shock propagates along the
$x$-axis, while the initial regular magnetic field is in the
$x$--$z$ plane. We show the $B_{y}$ and $B_{z}$ dependence on
$x$ in the left and right panels respectively.} \label{hyb_B}
\end{center}
\end{figure}

\begin{figure}[!ht]
\begin{tabular}{c}
\includegraphics[width=\textwidth,clip]{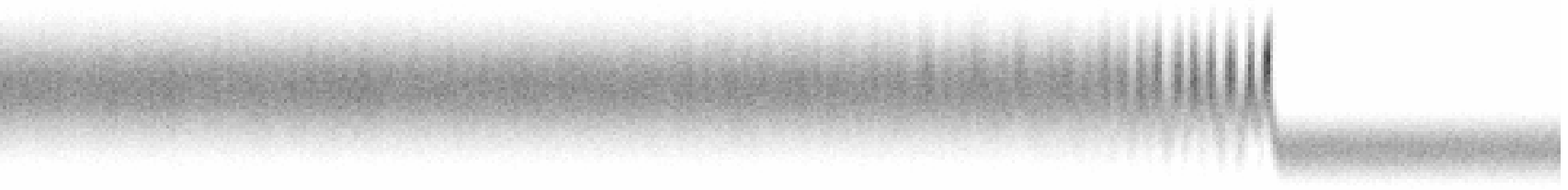}\\
\includegraphics[width=\textwidth,angle=0,clip]{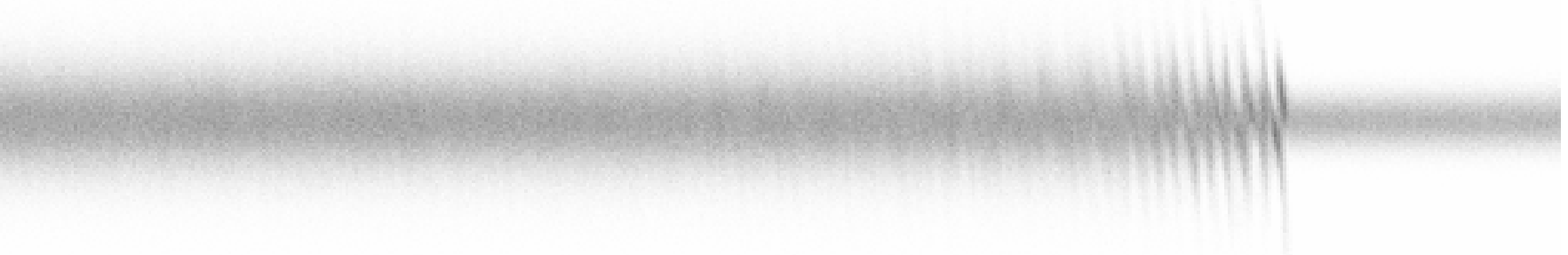} \\
\includegraphics[width=\textwidth,angle=0,clip]{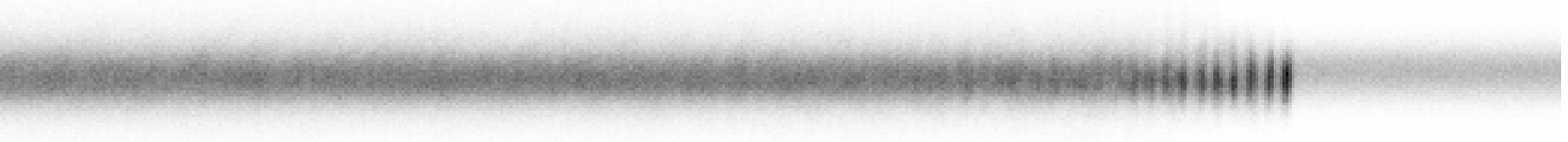} \\
\end{tabular}
\caption{Hybrid simulated proton phase density in a
quasi-perpendicular shock (80$\degmark$ inclination). The shock is
moving from left to right in the reference frame where the particle
reflecting wall is at rest. The figures show the proton phase
densities in ${v_x - x}$, ${v_y - x}$ and ${v_z - x}$ projections
from top to bottom respectively.}
 \label{hyb_phase}
\end{figure}

In Fig.~\ref{hyb_B} we show the structure of the magnetic field in a
quasi-perpendicular shock (inclination angle $\theta_{Bn} \approx$
80$\degmark$) simulated with a hybrid code for the upstream plasma
parameter $\beta \sim$ 1. The parameter $\beta = {\cal M}^2_{\rm
a}/{\cal M}^2_{\rm s}$, characterises the ratio of the thermal and
magnetic pressures. The shock is propagating along the $x$-axis
from the left to the right. The magnetic field is in the $x-z$
plane. The system is periodic in the $y$-dimension. Phase densities
of protons {$v_x - x$, $v_y - x$, $v_z - x$} are shown in
Fig.~\ref{hyb_phase} in the reference frame where the particle
reflecting wall (at far left) is at rest while the shock front is
moving. The incoming plasma beam in the simulation was composed of
protons (90~\%), alpha particles (9.9~\%) and a dynamically
insignificant fraction of oxygen ions (\ion{O}{vii}).

In most of the cases non-relativistic shocks simulated with
different hybrid codes had the upstream plasma parameter $\beta
\sim$ 1. In some cosmological shocks, for example in hot X-ray
clusters, plasma parameter $\beta$ could be $\sim$ 100 (see  \citealt{bykov2008} - Chapter 8,  this volume). The nature of collisionless
shocks in the hot low magnetised plasmas could be different from
that in case of $\beta \sim$ 1. While the processes of shock
formation in the high beta plasmas still require a careful study,
there are two experimental studies establishing the existence of the
collisionless shocks for that case. A collisionless shock in a
laboratory experiment with a laser-produced ablating plasma of
$\beta \sim$ 300 was found to have a thickness about 10
$c/\omega_{\rm pe}$, that is orders of magnitude less than the
Coulomb mean free path of both ions and electrons in that system
(see \citealt{Bell_ea88}). In space plasma the terrestrial bow shock
under high beta conditions was observed with the {\sl ISEE\,1} and
{\sl ISEE\,2} spacecraft by \citet{Farris_ea92}. These measurements
were compared with and found to be in agreement with the predicted
values of the Rankine-Hugoniot relations using the simple adiabatic
approximation and a ratio of specific heats, gamma, of 5/3. Large
magnetic field and density fluctuations were observed, but average
downstream plasma conditions well away from the shock were
relatively steady, near the predicted Rankine-Hugoniot values. The
magnetic disturbances persisted well downstream and a hot, dense ion
beam was detected leaking from the downstream region of the shock.

\subsection{Heating of ions in collisionless shocks}

 The heating processes in collisionless shocks are non-trivial. The
{\sl irreversible} transformation of a part of the kinetic energy
of the ordered bulk motion of the upstream flow into the energy of
the random motions of plasma particles in the downstream flow in
collisional shocks is due to Coulomb or atomic particle
collisions. In collisional non-radiative shocks without slowly
relaxing molecular components, the standard single-fluid RH
relations are applicable just after a few collisional lengths. The
standard single-component shock model predicts a particle
temperature ${\mathrm k}T = (3/16) mv_{\rm sh}^2$ for $\gamma = 5/3$.

The particle distributions in the collisionless shocks are not
Maxwellian. Thus, instead of the standard equilibrium temperature
the appropriate moments of the particle distribution function
characterising the width of randomised velocity distributions are
used. Moreover, the particle velocity distributions are typically
anisotropic. It is clearly seen in hybrid simulated proton phase
density:  in Fig.~\ref{hyb_phase} the velocity distribution widths
are different for different projections. One can see in
Fig.~\ref{hyb_phase} that a fraction of the incoming ions is reflected
by the shock magnetic field jump  providing multiple inter-penetrating flows of
gyrating ions. Then the field fluctuations randomise the ion phases
producing a "coarse-grained" distribution characterised by an
effective temperature estimated as the second moment of the velocity
distribution.

An analysis of interplanetary collisionless shock observations made
with {\sl Advanced Composition Explorer} by \citet{Korreck_ea07}
indicated that quasi-perpendicular shocks are heating ions more
efficiently than quasi-parallel shocks. It was also found that
effective temperatures of different ions are not necessarily
proportional to ion mass, but also depend on the shock inclination
angle and plasma parameter $\beta$. The widths of the collisionless
shocks are extremely narrow (below the astronomical unit) and thus,
the observed temperatures would depend on the temperature
equilibration processes (both Coulomb and collective) that we will
discuss elsewhere (e.g. \citealt{bykov2008} - Chapter 8,  this
volume). The temperature equilibration of different plasma
constituents in the WHIM can be studied with spatially resolved
spectroscopic observations and thus is a good test for shock models.

\subsection{Heating of electrons in collisionless shocks}

\begin{figure}    
\begin{center}
\includegraphics[width=\textwidth]{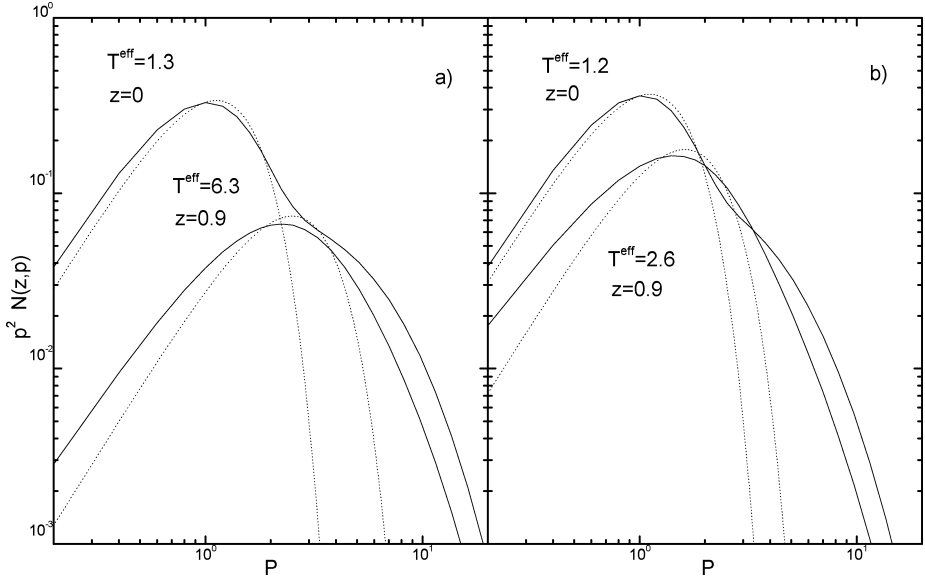}
\caption{Electron distribution function simulated in the model of
electron heating by strong ion gyroradii scale magnetic fluctuations
in a collisionless shock by \citet{BykovU99}. 
The left panel is for $\alpha=<(\delta B/B)^2> = 0.4$, 
the right panel for $\alpha=0.1$.} \label{el_temp}
\end{center}
\end{figure}

Electron kinetics in collisionless shocks are different from those of
ions. Since most of the observable emission comes from the
electrons, they require a careful study. Shocks transfer a fraction
of the bulk  kinetic energy of the ion flow into large amplitude
nonlinear magnetic fluctuations on a short scale of the transition
region (see Fig.~\ref{hyb_B}). It is important that the thermal
electron velocities in the ambient medium are higher than the shock
speed for a shock Mach number
 ${\cal M}_{\rm s}<\sqrt{m_{\rm p}/m_{\rm e}}$,
allowing for a nearly-isotropic angular distribution of the
electrons. Non-resonant interactions of these electrons with the
large-amplitude turbulent fluctuations in the shock transition
region  could result in collisionless heating and pre-acceleration
of the electrons \citep{BykovU99}. In Fig.~\ref{el_temp} a
simulated electron distribution ($p^2 N(z,p)$) is shown as a
function of the dimensionless electron momentum $p/\sqrt{2m_{\mathrm e}T_1}$,
where $T_{1}$ is the initial electron temperature in the far
upstream ($z \to -\infty$). The solid curves are the simulated
electron distribution functions at the left boundary ($\tilde{z}
=0$) of the transition region clearly seen in Fig.~\ref{hyb_B}, and
at the end of the region ($\tilde{z} =1$). Dotted lines are the
Maxwellian distribution fits allowing to estimate the effective
electron temperatures $T^{\mathrm{eff}}$ measured relative to  $T_1$. Note that
$T^{\mathrm{eff}} = 1.2$ at $\tilde{z} =0$ because of the electron
diffusivity effect. One may also clearly see the appearance of
non-thermal tails indicating a Fermi type acceleration. It is worth
noting that the presence of large-amplitude waves in the shock
transition region
 erodes many of the differences between quasi-parallel and
perpendicular shocks, making the electron injection mechanism in
that model to be similar for these shocks.

The analysis of observational data on both interplanetary and
supernova shocks by \citet{Ghavamian_ea07} indicates that theelectron heating efficiency i.e. $T_{\rm e}/T_{\rm i}$ is a
declining function of the shock velocity. These authors discussed a
model of electron heating with a constant level of electron heating
over a wide range of shock speeds (see also Fig.~4 in
\citealt{BykovU99}), while the ion heating is an increasing function
of shock velocity .

\subsection{Gas heating and entropy production in weak internal shocks}

Heating/acceleration efficiency with weak and moderate strength MHD
shocks can be estimated by calculating the energy dissipation rate
${\dot \varepsilon}_{\rm h}$ of a directed gas motion per unit area
of a weak shock. Defining ${\dot \varepsilon}_{\rm h} = v_{\rm sh}
\rho T \Delta s$, where $\Delta s$ is the difference of the
entropies (per unit mass) behind and ahead of the shock front, one
can evaluate the thermal gas heating as 
\be 
{\dot \varepsilon}_{\rm
h} = (5/4)({\cal M}_{\rm s} -1)^3 v_{\rm sh} \varepsilon_{\rm T},
\label{eq:wsh1} 
\ee 
where $\varepsilon_T$ is the gas internal energy
per unit volume (cf. Eq.~\ref{eq:entr}). The energy transfer to the
reflected nonthermal particles can be estimated from: 
\be {\dot
\varepsilon}_{\rm CR} = ({\cal M}_{\rm s} -1)^2 v_{\rm sh}
\varepsilon_{\rm B}, \label{eq:wsh2} 
\ee 
where $\varepsilon_B$ is
the magnetic energy density \citep{BykovT87}. From
Eq.~\ref{eq:wsh1} and Eq.~\ref{eq:wsh2} one can see that the gas
heating is of third order in $({\cal M}_{\rm s} -1) \ll 1$ (cf.
\citealt{LandauL59}), while the wave damping due to the particle
acceleration is of second order. Note that in the outer parts of
galaxy clusters one would typically expect $\varepsilon_{\rm T} \gg
\varepsilon_{\rm B}$. However,  the central regions of such a
cluster could have $\varepsilon_{\rm T}$ comparable to
$\varepsilon_{\rm B}$, as it is the case in the Milky Way. Thus, the
weak shocks in the central regions could efficiently accelerate
nonthermal particles, reducing the heating of the gas. Particle
acceleration by an ensemble of large scale shocks in a cluster of
galaxies can create a population of non-thermal particles of
sizeable pressure. This may imply a non-steady evolution of
non-thermal pressure as modelled by \citet{Bykov01}.

\section{Energetic particle acceleration in collisionless shocks}
\label{CR}

 The reflected ions with a gyro-radius exceeding the width
of the shock transition region can then be efficiently accelerated,
via the Fermi mechanism, by converging plasma flows carrying
magnetic inhomogeneities and MHD waves. In perpendicular shock a net
transverse particle momentum gain is due to the work of the electric
field  on the particle drift motion. The electric field
perpendicular to the shock normal exists in all the reference frames
for the perpendicular shock. The particle of a momentum $\mathbf{p}$
crossing back and forth the shock front and being scattered by MHD
waves carried with a flow of velocity $\mathbf{u}$ would undergo a
momentum increment $\Delta p \approx
\mathbf{p}\cdot\frac{\mathbf{u}}{v} + O((u/v)^2)$ per scattering.  A
velocity profile in the plane shock is illustrated by the dashed
line in Fig.~\ref{sketch} in the test particle case where one
neglects the back reaction effect of accelerated particles on the
shock. One way to calculate the accelerated particle spectra in a
scattering medium is to use the kinetic equation in the diffusion
approximation. The scattering medium in that approach is
characterised by momentum dependent particle diffusion coefficients
$k_1(p)$ and $k_2(p)$. The shock is considered as a bulk velocity
jump (see the dashed line in Fig.~\ref{sketch}) assuming that the
test particles are injected at $p = p_0$ and the gyroradii of the
particles are larger than the shock width. Therefore, in a test
particle case particles must be injected at some super-thermal
energy to be accelerated by the shock. A solution to the kinetic
equation for a nearly isotropic test particle distribution in the
phase space is a power-law momentum distribution $f(p,x) \propto
(p/p_0)^{-b}$, $p\geq p_0 $ where the index
 \be
 b =
\frac{3r}{r-1} \ee 
depends on the shock compression ratio $r$
\citep{Axford_ea77,Krimskii77,Bell78a,BlandfordO78}. The CR spatial distribution in the model is
illustrated in Fig.~\ref{sketch}. For a strong shock of ${\cal
M}_{\rm s}\gg 1$ and ${\cal M}_{\rm a}\gg 1$ the compression ratio
given by Eqs.~\ref{eq:rhr} and \ref{eq:rhrp} is close to 4 if $\gamma_{\mathrm g}
= 5/3$ (or even larger if relativistic gas dominates the equation of
state). The pressure of the accelerated particles is
\begin{equation}\label{rel_pres}
P_{\mathrm{CR}} = \frac{4 \pi}{3} \int\limits_{p_0}^{\infty}\, p\,v\,f(p,x)\,p^2\,
{\mathrm d}p.
\end{equation}
Then for $b=4$ one may see that  $P_{\mathrm{CR}} \propto \ln(p_{\max}/p_0)$
indicating a potentially large cosmic ray (CR) pressure, if CRs are
accelerated to $p_{\max} \gg p_0$. The maximal energy of accelerated
test particles depends on the diffusion coefficients, bulk velocity
and scale-size of the system. The finite scale-size of the shock is
usually accounted for by an energy dependent free escape boundary
located either in the upstream or in the downstream. For electrons
$p_{\max}$ can also be limited by synchrotron (or inverse-Compton)
losses of relativistic particles.

The test particle shock acceleration time $\tau_{\mathrm a}(p)$
 can be estimated from the equation
\begin{equation}\label{tacc}
\tau_{\rm a}(p) = \frac{3}{u_{1n} - u_{2n}} \int\limits_{p_0}^{p} \left(
\frac{k_1(p)}{u_{1n}} + \frac{k_{2}(p)}{u_{2n}}\right)\frac{{\mathrm d}p}{p}
\end{equation}
where the normal components of the shock upstream and downstream
bulk velocities $u_{1n}, u_{2n}$ are measured in the shock rest
frame. Estimations based on a more rigorous approach distinguishing
between the mean acceleration time and the variance does not change
the results substantially, given the uncertainties in the diffusion
model.

\section{Cosmic-ray modified multifluid shocks}
\label{CRS}

\begin{figure}    
\begin{center}
\hbox{
\psfig{file=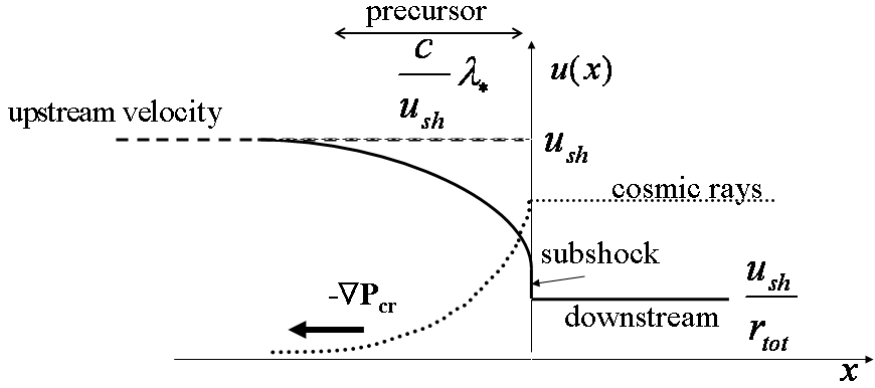,width=11.5cm,clip=}}
\caption{A sketch illustrating the structure of a cosmic-ray
modified shock. The dashed line is the shock velocity jump
corresponding to the test particle case. The dotted line is a
spatial distribution of accelerated particles at some momentum $p
\gg p_0$. The solid line is the CR modified shock velocity profile with
the precursor and subshock indicated.} \label{sketch}
\end{center}
\end{figure}

The efficiency of the upstream plasma flow energy conversion into
nonthermal particles could be high enough providing a hard spectrum
of nonthermal particles up to some maximal energy
${\varepsilon}_{\star}$. If the efficiency of ram energy transfer to
the energetic particles is high enough, an extended shock precursor
appears due to the incoming plasma flow deceleration by the fast
particle pressure. The precursor scale $L$ is of the order of
$(c/v_{\rm sh})  \lambda_{\star}$ -- orders of magnitude larger
than the width of the shock transition region (see
Fig.~\ref{sketch}). Here $\lambda_{\star}$ is the maximal mean free
path of a particle in the energy-containing part of the spectrum and
$v_{\rm sh}$ is the shock velocity. We shall later refer to these
energetic particles as cosmic rays.

It has been shown that the front of a strong collisionless shock
wave consists of an extended precursor and a viscous velocity
discontinuity (subshock) of a local Mach number that is smaller than
the total Mach number of the shock wave (see Fig.~\ref{sketch}). The
compression of matter at the subshock can be much lower than the
total compression of the medium in the shock wave with allowance for
high compression in the precursor. We shall refer later to such shocks
as CR-modified.

The large scale ("macroscopic") structure of a CR-modified shock can
be modelled by a two-fluid approach with a kinetic description of
nonthermal particles (see e.g. \citealt{BlandfordE87,Berezhko_ea96,MalkovD01,Blasi04} and
references therein) or by a Monte Carlo method (e.g.
\citealt{JonesE91,Ellison_ea96}). In both methods some
suitable parameterisation of particle scattering process must be
postulated {\it a priori}. Monte-Carlo simulations, however, have no
assumption of isotropy for particle distributions, and that allows
an internally self-consistent treatment of thermal particle
injection. While the injection depends on the assumptions made for
the particle pitch-angle scattering, these assumptions are applied
equally to particles of all energies. The Monte Carlo technique
eliminates a free injection parameter, which is present in all
models based on the diffusion approximation and is used to set the
injection efficiency. The strong feedback between injection, shock
structure, and magnetic field amplification makes this property of
the Monte Carlo technique particularly important. The Monte Carlo
technique allows to iteratively obtain a shock velocity profile and
particle distribution function conserving mass, momentum and energy
fluxes taking into account the nonlinear feedback from the
accelerated energetic particles.

\begin{figure}    
\begin{center}
\hbox{
\includegraphics[width=0.49\textwidth,height=6cm]{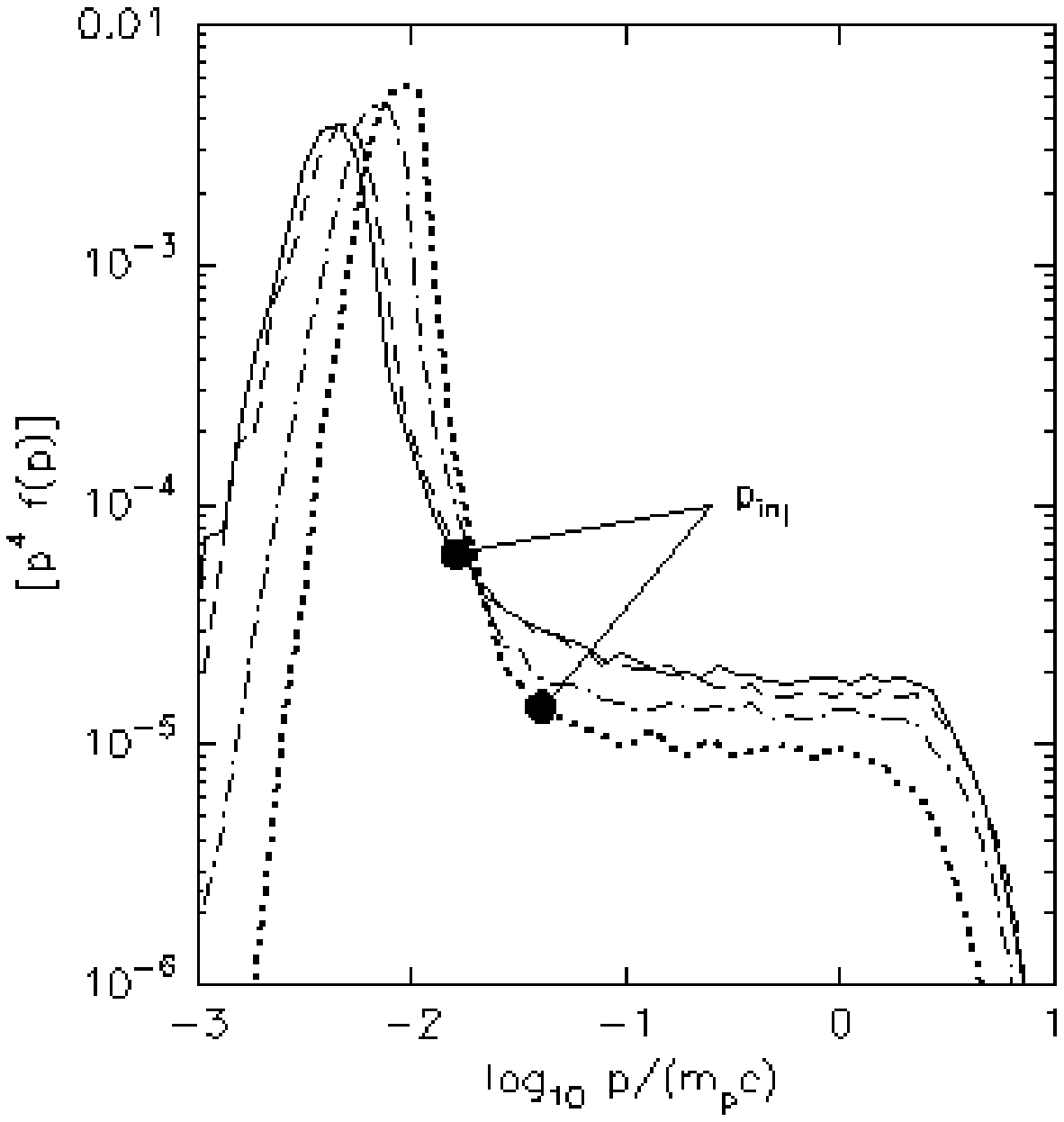}
\includegraphics[width=0.49\textwidth,height=6cm]{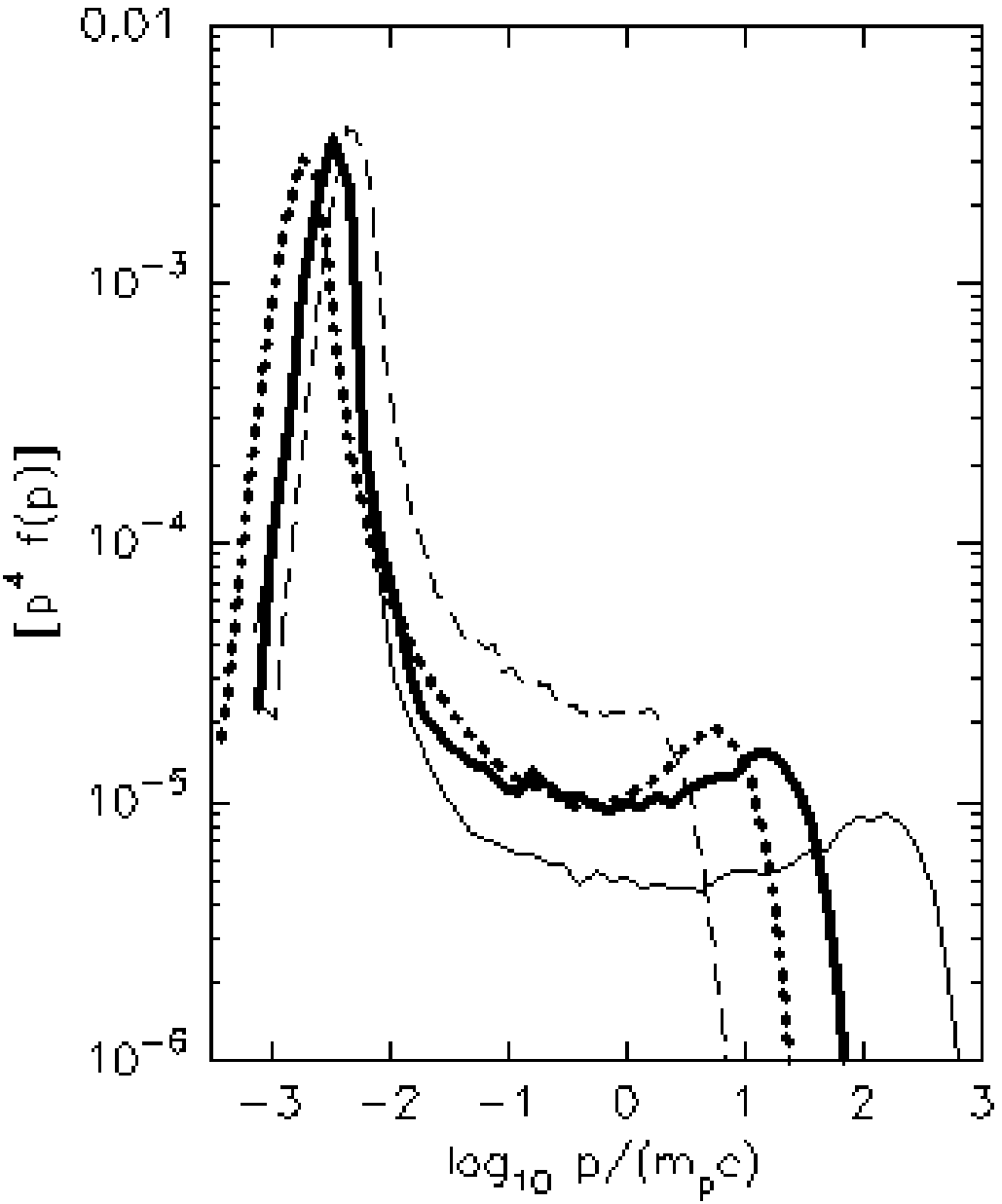}
}
\caption{Spectra of protons accelerated by a strong shock. The
spectra were simulated with a non-linear Monte-Carlo model which
accounts for particle injection and magnetic field amplification by
the shock (for details see \protect\citealt{Vladimirov_ea06}). On the left
panel the transitions from the thermal-like component to the high
energy tail are marked by $p_{\mathrm{inj}}$. On the right panel different
curves correspond to different locations of the free escape boundary
(see in the text).} \label{fp}
\end{center}
\end{figure}

In Fig.~\ref{fp} Monte Carlo simulated proton spectra (multiplied by
$[p/(m_{\rm p}c)]^4$) are shown, in the downstream shock from
\citet{Vladimirov_ea06}. To illustrate the dependence of the maximal
energy of an accelerated proton on the system scale size, a free
escape boundary condition was applied at some distance from the
subshock position in the shock rest frame. The heavy solid and
dotted curves  in the right panel correspond to the free escape
boundary located at a distance 10$^4r_{\rm g1}$ (where $r_{\rm g1}= m_{\mathrm p}
v_{\rm sh} {\mathrm c}/{\mathrm e} B_1$), the dashed curve has 10$^3r_{\rm g1}$, and the
light solid curve has 10$^5r_{\rm g1}$. The simulations were done for
a supernova shock in the interstellar medium with a shock speed
$v_{\rm sh} = 5000 $~km\,s$^{-1}$ and an unshocked proton number density $n_1 =
1$~cm$^{-3}$. In the left panel the spectra are given for the same
position of the free escape boundary, but for different
prescriptions of the scattering model.

\subsection{Magnetic field amplification in CR-dominated shocks}

An important predicted feature of strong shocks with efficient CR
acceleration is the possibility to amplify an initial seed magnetic
field by orders of magnitude (e.g. \citealt{BellL01,Bell04}).
CR current and CR pressure gradient upstream of the strong shock
could drive magnetic fluctuations on the shock precursor scale
length. The CR-shock precursor scale $L$ is $\sim (c/v_{\rm sh})
 \lambda_{\star}$ which is expected to be above a kpc,
moreover, the width is $L \gtrsim$ 100 kpc for a shock of  a size
comparable to that of a galaxy cluster. The precursor scale size $L$
is $\gg$ 10$^9$ times larger than the subshock transition region
where strong small scale magnetic field fluctuations are directly
produced by instabilities of super-Alfv\'enic bulk plasma flows
illustrated in Fig.~\ref{hyb_B}. That small scale fluctuations are
responsible for bulk plasma motion dissipation process and adiabatic
amplification of the transverse magnetic field in collisionless
shocks. At the same time the collisionless dissipation process is
thought to inject a minor fraction of incoming particles to be
accelerated to high energies by Fermi mechanism. Recent models of
diffusive shock acceleration allows a substantial fraction (say,
30~\%) of the MHD shock ram pressure to be converted to accelerated
particles filling a vicinity of the shock of the scale $L$. The large
scale current and density gradient of the accelerated CRs may
convert a fraction of the CR energy to magnetic field due to
multifluid instabilities of different kinds providing a way to
amplify the initial magnetic field by a factor  larger than the
shock compression ratio.

Recent non-linear simulations of magnetic field amplification in
diffusive shock acceleration by a Monte-Carlo model
\citep{Vladimirov_ea06} and a kinetic model \citep{AmatoB06}
confirmed the possibility of a significant effect. The amplitude of
the fluctuating magnetic field energy density W$_{\rm B}$ is of the
order of the shock accelerated CR pressure which is in turn a
substantial fraction of the shock ram pressure $0.5\, \rho_1 \,
v_{\rm sh}^2$. Here $\rho_1$ is the shock upstream ambient gas
density.

For  typical cluster parameters the discussed mechanism could
provide a $\mu$G range magnetic field amplitude in a hundred kpc
range scale of CR-modified shock precursor. The Faraday rotation
measure $RM$ provided by a strong CR-dominated shock in a cluster
can reach values of $\gtrsim$ 10 rad m$^{-2}$ and even a few times
higher. For the case of the so-called Bohm diffusion model the
rotation measure $RM$ is proportional to the maximal energy of the
ions in the energy-containing part of the CR-spectrum accelerated by
the shock. Radio observations, Faraday rotation and
synchrotron-Compton emission measurements are used to estimate the
magnetic fields in clusters (e.g. \citealt{CarilliT02,Newman_ea02}). Large filaments of  polarised radio emission
of scale size about 400 kpc were discovered by \citet{Govoni_ea05}
in the halo of the cluster of galaxies Abell~2255 and by
\citet{Bagchi_ea06} in Abell~3376 (see Fig.~\ref{A3376_radio}). They
could be connected to large scale shocks due to accretion/merging
activity of the cluster.

\subsection{Gas heating and entropy production in strong CR-modified
shocks}

An exact modelling of a collisionless shock structure taking into account
the nonthermal particle acceleration effect requires the
nonperturbative self-consistent description of {\sl a
multi-component and multi-scale system} including strong
MHD-turbulence dynamics. Such a modelling is not feasible at the
moment. Instead, a simplified description of a multi-fluid strong
shock structure can be used with an appropriate parameterisation of the
extended pre-shock and of the gas subshock. The predicted observable
characteristics of the shocks can be confronted to the
observational data. We will now discuss  the effects of plasma
heating by modified shocks and then make some specific predictions for
possible observational tests.

In the shocks with efficient high energy particle acceleration  the
energy flux carried away by escaping energetic particles $Q_{\rm
esc}$  must be accounted for in the energy continuity equations. The
energy loss results in a  lower effective adiabatic index, but it
allows to increase the total compression of the gas in the shock
downstream.

\begin{figure}    
\begin{center}

\includegraphics[width=0.6\textwidth]{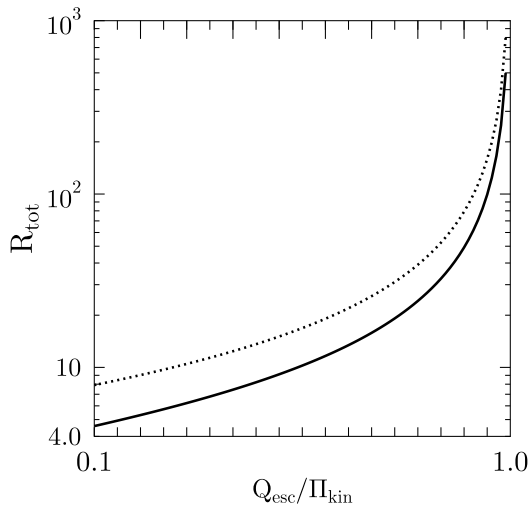}
\caption{Total compression ratio $r_{\rm tot}$ of a strong MHD shock
modified by efficient particle acceleration as a function of the
energy escape flux $Q_{\rm esc}/\Pi_{\rm kin}$ carried by energetic
particles, where $\Pi_{\rm kin} = \rho_1 v_{\rm sh}^3/2$. The upper
curve (dotted) corresponds to an effective adiabatic exponent
$\gamma$ = 4/3 (relativistic gas), while the lower (solid) curve
corresponds to $\gamma$ = 5/3.} \label{compr1}
\end{center}
\end{figure}

\begin{figure}
\begin{center}
\psfig{file=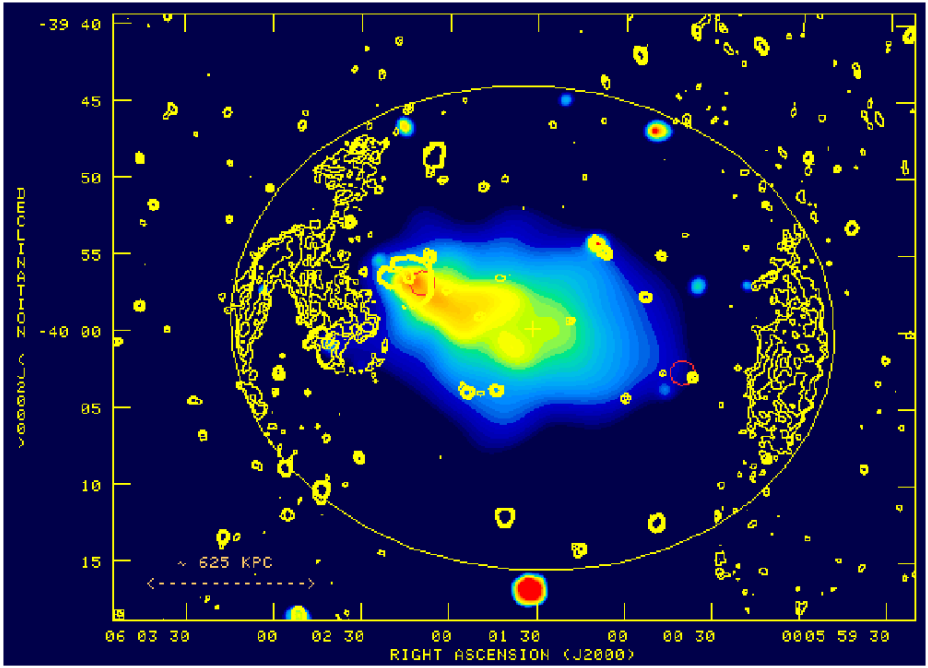,width=9.cm,angle=-90} \caption{Abell~3376
(taken from \protect\citet{Bagchi_ea06}: X-ray emission from XMM-Newton
archive data, with VLA 1.4~GHz radio contours superimposed.  The
ellipse shows an elliptical fit to the peripheral radio structures,
and the `+' marks its centre.
The 
circles mark the positions of the two brightest cluster
galaxies. } \label{A3376_radio}
\end{center}
\end{figure}

The total compression ratio $r_{\rm tot}$ of a strong MHD shock
modified by an efficient nonthermal particle acceleration can be
estimated as
\begin{equation}
r_{\rm tot} = \frac{\gamma + 1}{ \gamma - \sqrt{1 + 2(\gamma^2
-1)Q_{\rm esc}/\rho_a v_{\rm sh}^3}}, \label{eq:crrt}
\end{equation}
assuming that the energy density in the shock upstream  is dominated
by the ram pressure and that the CR escape is through the cut-off
momentum regime (e.g. \citealt{MalkovD01}). Here $\gamma$ is the
effective adiabatic exponent. In Fig.~\ref{compr1} we illustrate the
dependence of the compression ratio on $Q_{\rm esc}/\rho_a v_{\rm
sh}^3$ for  $\gamma$ = 4/3 and 5/3 assuming that the effective
adiabatic exponent is between the two values depending on the
spectrum of the accelerated relativistic particles.

The distribution function of nonthermal particles and the bulk flow
profile in the shock upstream region are sensitive to the total
compression ratio $r_{\rm tot}$. Thus, the exact calculation of the
escape flux $Q_{\rm esc}$ can be performed only in fully nonlinear
kinetic simulations. Nevertheless, an approximate iterative approach
(e.g. in the Monte Carlo model discussed above) can be used to make
the steady-state distribution function consistent with the shock
compression {\sl assuming some diffusion model}. The subshock is the
standard gas viscous shock of a Mach number ${\cal M}_{\rm sub}$.
For that simplified {\sl two-fluid} model of a strong CR-modified
shock the effective ion temperature in the downstream $T^{(2)}_{\rm
i}$ can be estimated for a shock of a given velocity, if $r_{\rm
tot}$ and $r_{\rm sub}$ are known:
\begin{equation}
T^{(2)}_{\rm i} \approx \phi({\cal M}_{\rm sub})  \  \frac{ \mu~
v_{\rm sh}^2}{\gamma_{\rm g}r_{\rm tot}^2(v_{\rm sh})} ,~~{\rm
where}~~ \phi({\cal M}_{\rm sub}) = \frac{2 \gamma_{\rm g} {\cal
M}_{\rm sub}^2 - (\gamma_{\rm g} -1)}{(\gamma_{\rm g} -1){\cal
M}_{\rm sub}^2 + 2}. \label{eq:tcr}
\end{equation}

Single fluid strong shock heating represents the limit ${\cal
M}_{\rm sub} = {\cal M}_{\rm s} \gg 1$, since there is no precursor
in that case, resulting in Eq.~\ref{eq:rht1}. In single-fluid
systems the compression ratio $r_{\rm tot} = r_{\rm sub} \rightarrow
(\gamma_{\rm g} +1)/(\gamma_{\rm g} -1)$ does not depend on the
shock velocity and Eq.~\ref{eq:tcr} reduces to Eq.~\ref{eq:rht1}.
However, in multi-fluid shocks the total compression ratio depends
on the shock velocity and could be substantially higher than that in
the single-fluid case. This implies somewhat lower postshock ion
temperatures for the strong multi-fluid shock of the same velocity
and could be tested observationally. It is convenient to introduce
the scaling $r_{\rm tot}(v_{\rm sh}) \propto v_{\rm sh}^{\xi}$ to
describe the different cases of strong shock heating (see
\citet{Bykov05} for details). Then from Eq.~\ref{eq:tcr}, $T_{\rm i2}
\propto \phi({\cal M}_{\mathrm sub})\  v_{\rm sh}^{2(1-\xi)}$. The
subshock Mach number ${\cal M}_{\rm sub}$ depends, in general, on
${\cal M}_{\rm s}$ and ${\cal M}_{\rm a}$. Thus, an  index $\sigma$
approximates the velocity dependence of $\phi({\cal M}_{\rm sub})
\propto v_{\rm sh}^{\sigma}$. Finally, if $T_{\rm i2} \propto v_{\rm
sh}^a$, then the index $a = 2(1-\xi) + \sigma$ . For the case of
shock precursor heating by CR generated \alf waves, the index $a
\approx$  1.25 \citep{Bykov05}.

A distinctive feature of multi-fluid shocks is their high gas
compression $r_{\rm tot}(v_{\rm sh})$ that could be well above the
single fluid shock limit $(\gamma_{\rm g} +1)/(\gamma_{\rm g}-1)$
(see Fig.~\ref{compr1}). At the same time entropy production for a
strong multi-fluid shock scales as $r_{\rm tot}(v_{\rm
sh})^{-(\gamma_{\rm g} +1)}$ and it is significantly reduced
compared to the single-fluid shock of the same velocity. The effects
are due to energetic particle acceleration and magnetic field
amplification.

Energetic particles penetrate into the shock upstream region. They
are coupled with the upstream gas through fluctuating magnetic
fields (including the \alf waves generated by the energetic
particles). Magnetic field dissipation provides gas preheating and
entropy production in the extended shock precursor. Such a heated
pre-shock region of ${\mathrm k}T\lesssim 0.5$~keV would appear as an extended
filament of width $L \sim (c/v_{\rm sh})  \lambda_{\star} \gtrsim
3\times 10^{14}  \epsilon_{\star}  B_{-6}^{-1}$ cm. Here
$\epsilon_{\star}$ (in GeV) is the highest energy of the hard branch
of the accelerated particle spectrum. If $B_{-6} \sim$ 0.1 in the
cluster outskirts and if the hard spectrum of energetic nuclei
extends to $\sim 10^9$ GeV (cf. \citealt{Norman_ea95}) we have $L
\sim 1$~Mpc and even wider. Projected on a hot X-ray cluster, such
filaments could produce a soft X-ray component "excessive" to that
produced by the hot cluster. A warm gas ($\sim$ 0.2 keV) emission
filament found with \xmm\ in the outskirts of the Coma cluster by
\citet{Finoguenov_ea03} could be an extended heated precursor of a
strong multi-fluid accretion shock. For a detailed review of the
soft X-ray/EUV excesses see \citealt{durret2008} - Chapter 4, this
volume.

\subsection{ICM entropy production by multifluid accretion shocks }

Cold gas falling into the dark matter (DM) dominated gravitational
well passes through a strong accretion shock. The shock is a source
of gas entropy production in the intercluster medium (ICM) (e.g.
\citealt{KnightP97,TozziN01,Voit_ea03}). The
post-shock entropy $K = K_{\rm b}\, T/\rho^{2/3}$ used in the ICM
analysis and simulations (e.g. \citealt{Bialek_ea01}) is related to
the standard thermodynamic entropy $s$ through $K \propto
\exp(s/c_{\mathrm v})$. In the standard scenario with a single-fluid accretion
shock the post-shock entropy  scales $K_{\rm sf} \propto v_{\rm
sh}^2  \rho_1^{-2/3}$ (e.g. \citealt{Voit_ea03}).

The multi-fluid nature of the collisionless accretion shock modifies
the standard scaling relation to be
\begin{equation}
K_{\rm mf} \propto v_{\rm sh}^2 [r_{\rm tot}(v_{\rm sh})]^{-(1
+\gamma_{\rm g})} \phi({\cal M}_{\mathrm{sub}})\, 
\rho_1^{(1-\gamma_{\rm g})}.
\end{equation}
The compression ratio in CR-shocks is higher than in a strong
single-fluid shock of the same velocity resulting in reduced
post-shock entropy production. For example, in the case of \alf
heating the post-shock entropy of a multi-fluid shock reduces as
$K_{\rm mf}/K_{\rm sf} \sim (15/{\cal M}_{\rm a})$ for ${\cal
M}_{\rm a} > 15$ and ${\cal M}_{\rm s}^2 > {\cal M}_{\rm a}$. Here
and below in numerical estimations we assume $\gamma_{\rm g}=5/3$,
though a non-thermal baryonic component could reduce the index
$\gamma_{\rm g}$.

Since $r_{\rm tot}(v_{\rm sh})$ and $\phi({\cal M}_{\rm sub})$ are
shock velocity dependent, the simple scaling $K \propto v_{\rm sh}^2
 \rho_1^{-2/3}$ is not valid.  In CR-modified shocks $K_{\rm
mf} \propto v_{\rm sh}^{\nu}  \rho_1^{(1-\gamma_{\rm g})}$ or
$K_{\rm mf} \propto T^{\nu/a}$, where $\nu =2 - (1+\gamma_{\rm
g}) \xi +\sigma$. For the case of  \alf wave heating the index
$\nu$ is $\lesssim 1.25$ and $K_{\rm mf}$ is $\propto T^{0.8}$ assuming
$\gamma_{\rm g} = 5/3$. Recently \citet{Ponman_ea03} and
\citet{Piffaretti_ea05} found that the dispersion in the observed
cluster entropy profiles is smaller if an empirical relation $K
\propto T^{0.65}$ is used instead of the standard $K \propto T$ (see
also \citealt{Pratt_ea06}).

Consider the simple model of smooth accretion of cold gas through a
strong accretion shock by \citet{Voit_ea03}. The gas of velocity
$v_{\rm ac}$ accretes at a rate ${\dot M}_g$ through the shock at a
radius $r_{\rm ac}$ where
\begin{equation}
{\dot M}_g = 4\pi r_{\rm ac}^2 \rho_1 v_{\rm ac},~~~ v_{\rm
ac}^2=2GM\xi r_{\rm ac}^{-1},~~~\xi = 1 - r_{\rm ac}/r_{\rm ta}.
\label{eq:acr1}
\end{equation}
Here $M(t)$ is the cluster mass and $r_{\rm ta}$ is the matter
turnaround radius. Then the entropy $K_{\rm mf}$ just behind the
multi-fluid shock is expressed through $T^{(2)}_i(v_{\rm ac})$ and
$\rho_2 = r_{\rm tot}(v_{\rm ac}) \rho_1$. In the \alf wave
heating case $K_{\rm mf}(t)\propto (Mt)^{(1+\sigma)/3}$, instead of
$K_{\rm sf}(t) \propto (Mt)^{2/3}$ in the single-fluid regime. A
multi-fluid shock results in a slower post-shock entropy production.
As we have noted above, the regime of CR-shock compression depends
on the plasma parameter $\beta$ in the infalling gas. The plasma
parameter $\beta$ is currently poorly known because the intercluster
magnetic fields are not well constrained. The effects of shock
modifications are important for both the models of smooth accretion
of cold gas and for accretion of hierarchical structures.

Preheating of accreting gas by different physical processes (e.g.
due to early star formation in a protocluster region) was suggested
by \citet{EvrardH91}, as a possible reason for the breaking of the
scaling relations for pure gravitational cluster compression by
\citet{Kaiser86}. The observed high metallicity of clusters at
different redshifts indicates that strong starburst activity was
highly likely at some stage. The preheating produces some initial
level of gas entropy ("entropy floor", see e.g. extensive
simulations by \citealt{Bialek_ea01,Borgani_ea01,Borgani_ea05}). Multi-fluid strong shocks provide a natural
alternative way of preheating accreting gas. The non-thermal
components are essential for detailed modelling of global properties
of X-ray clusters, including the mass-temperature and
luminosity-temperature relations \citep{Ostriker_ea05}.

\begin{figure}
\begin{center}
\psfig{file=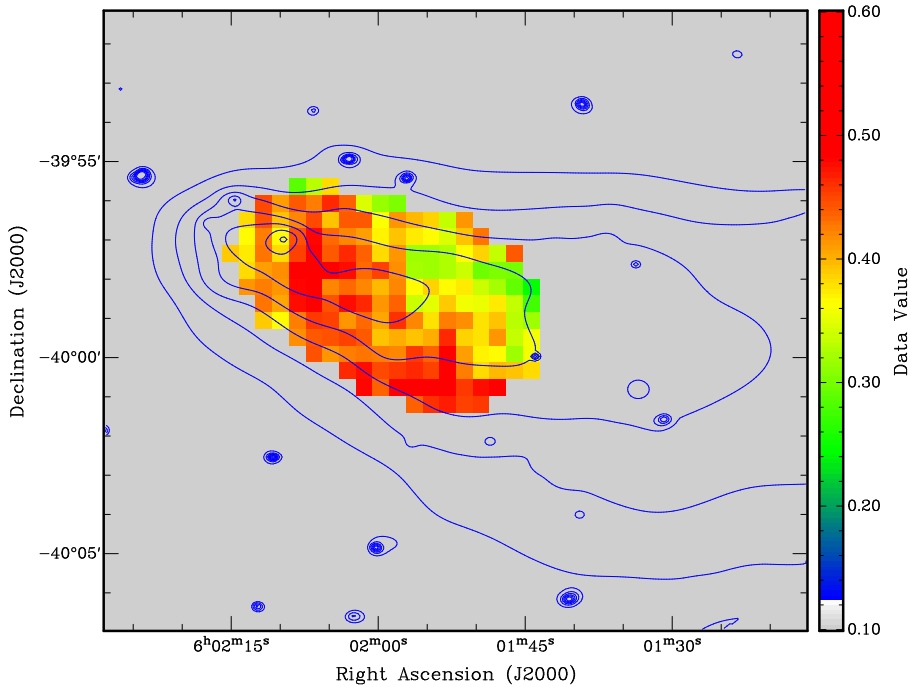,width=9.cm,angle=-90} \caption{Metallicity
map of the intracluster gas in Abell~3376 (taken from
\protect\citealt{Bagchi_ea06}).} \label{A3376_Z}
\end{center}
\end{figure}

\section{Shocks in large-scale structure}
\label{LSS}

Simulations of the cosmic large-scale structure (LSS) predict that
about 40$-$50~\% of baryons at epoch $z < 2$ could reside in the
Warm-Hot Intergalactic Medium (WHIM) with temperatures $10^5-10^7$~K at moderate overdensities $\delta \sim 10$ \citep{CenO99,Dave_ea01}. The WHIM heating is due to shocks driven by
gravitationally accelerated flows in the LSS structure formation
scenario (e.g. \citealt{Kang_ea07}). The simulations demonstrate that
the X-ray and ultraviolet \ion{O}{vi}, \ion{O}{vii} and \ion{O}{viii} lines and
the \ion{H}{i} Ly$\alpha$ line traces the low-density cosmic web
filamentary structures. Intervening metal absorption systems of
highly ionised C, N, O, Ne in the soft X-ray spectra of bright AGN
were suggested to trace the WHIM. The detection of shocked WHIM
requires very sensitive UV and X-ray detectors, both for absorption
and for emission processes (see e.g. \citealt{kaastra2008,richter2008} - Chapters 9 and 3, this volume). Dedicated future
missions like the
 {\sl Cosmic Origin Spectrograph} ({\sl COS}), the X-Ray Evolving
Universe Spectrometer ({\sl XEUS}), {\sl Constellation-X} and
  the Diffuse Intergalactic Oxygen Surveyor ({\sl DIOS}) will provide
  high resolution spectroscopy of the shocked WHIM. The WHIM ions of
  different charge states have highly non-equilibrium (anisotropic)
  initial states just behind a collisionless shock  that relaxes
  to equilibrium states through Coulomb collisions. As was discussed above
 a strong collisionless shock could generate a spectrum of
 MHD-fluctuations. These MHD-fluctuations can carry a substantial
 fraction of the shock ram pressure. The velocity fluctuations will
 result in non-thermal
broadening of the lines, potentially important for simulations of
emission/absorption spectra of the WHIM and observational data
analysis. Specific features  of collisionless shock heating of the
WHIM ions are discussed in
\citealt{bykov2008}  - Chapter 8,   this volume. In this paper we discuss only a few
observations of clusters of galaxies.

\subsection{Evidence for shocks in galaxy clusters}

Clusters of galaxies are believed to form within the hierarchical
build up of the large scale structure of the Universe. Small objects
collapse first and then merge in a complex manner to form larger and
larger structures. Therefore, once in a while during their
formation, clusters of galaxies undergo so called major merger
events. In such events, proto-cluster structures of similar masses
(typical ratios  1:10 $-$ 1:1) are colliding with super sonic
velocities (typically several 1000~km\,s$^{-1}$). These merging events are a
source of shocks and turbulence. They redistribute and amplify
magnetic fields, and they are a source of acceleration of
relativistic particles within the intracluster medium. Additionally,
it is expected that the accretion of the diffuse, unprocessed (and
therefore relatively cold) matter onto the DM node of the cosmic web
creates a virialisation shock (also called accretion shock), which
is expected to be located far in the cluster periphery (typically a
few Mpc from the cluster centre for massive systems). Some examples
of relevant cosmological simulations are presented by \citealt{dolag2008} - Chapter 15, this volume.

X-ray observations, revealing the thermodynamical state of the intracluster medium are therefore the natural means for searching for the
signatures of such non thermal phenomena. However, due to biases in
the observational processes, caused by the complex temperature
structure of the intracluster medium, such signatures are very hard
to detect (for a more detailed discussion see \citet{Mazzotta_ea04}
and references therein). Nevertheless, some detections of shocks in
galaxy clusters have been revealed by high resolution {\sl Chandra}
and {\sl XMM-Newton} observations. For a recent review see
\citet{MarkevitchV07}. One of the most spectacular examples of a
merging galaxy clusters is the case of 1E~0657$-$56
\citep{Markevitch_ea02}.

The sensitivities of current X-ray instruments are not sufficient to
map the state and structure of the intracluster medium in the
periphery of galaxy clusters. However, the discovery of arc like
radio emission in the periphery of some clusters (so called radio
relics), are thought to trace shocks running through the intracluster medium. 
Spectacular examples are Abell~3667 (see also \citealt{ferrari2008} - 
Chapter 6, this volume) or the radio relics in Abell~3376 recently 
discovered by \citet{Bagchi_ea06}, see Fig.~\ref{A3376_radio}. 
Thereby, such radio
observations are currently the only possibility to observe shocks
outside the central regions of galaxy clusters. A more detailed
discussion of numerical models of such radio relics is provided 
by \citealt{dolag2008} - Chapter 15, this volume.

Shocks are also expected to trigger star formation, as indicated by
numerical simulations (e.g. \citealt{Bekki99}), which will leave
detectable imprints in the intracluster medium even long (several
Gyr) after the shock passed through. Multiple supernova explosions
in the star forming regions (superbubbles) will additionally produce
copious small scale shocks and accelerate non-thermal particles (e.g.
\citealt{Bykov01}). One of the tracers for these processes can be excess
metallicity in the intracluster medium, produced by the enhanced
star formation period (e.g. \citealt{Schindler_ea05}).
Fig.~\ref{A3376_Z} shows the inferred metallicity map for
Abell~3376, indicating previous merger activity of the cluster (see
\citealt{Bagchi_ea06}).

\section{Summary}

Cosmological shocks convert a fraction of the energy of
gravitationally accelerated flows to internal energy of the gas.
They heat and compress the gas and can also accelerate energetic
non-thermal particles and amplify magnetic fields. We discussed some
specific features of cosmological shocks.

$\bullet$ The standard Rankine-Hugoniot relations based on the
conservation laws for a steady single-fluid MHD shock allow to
calculate the state of the fluid behind the shock once the upstream
state and the shock strength are known. The coplanarity theorem for
a plane ideal MHD shock states that the upstream and downstream bulk
velocities, magnetic fields and the shock normal all lie in the same
plane.

$\bullet$ Cosmological plasma shocks are likely to be collisionless
as many other astrophysical shocks observed in the heliosphere and
in supernova remnants.  We review the basic plasma processes
responsible for the microscopic structure of collisionless shocks.

$\bullet$ Collisionless shock heating of ions results in a
non-equilibrium  state just behind a very thin magnetic ramp region
with a strongly anisotropic quasi-Maxwellian ion distributions. The
possibility of collisionless heating of electrons by electromagnetic
fluctuations in the magnetic ramp region depends on the extension of
the fluctuation spectra to the electron gyro-scales, and could
depend on the shock Mach number. Then the Coulomb equilibration
processes are operating on the scales much larger than the
collisionless shock width.

$\bullet$ Extended MHD shock waves propagating in turbulent media
could accelerate energetic particles  both by Fermi type
acceleration in converging plasma flows and by DC electric field in
quasi-perpendicular shocks. If the acceleration is efficient, then
the strong shock could convert a substantial fraction (more than
10~\%) of the power dissipated by the upstream bulk flow to energetic
particles (cosmic rays). The compression ratio $r_{\rm tot}$ at such
a shock can be much higher, while the ion temperature behind the
shock $\propto r_{\rm tot}^{-2}$ and the post-shock entropy are
lower, than that in a standard single fluid shock. The shock
structure consists of an extended precursor and a viscous velocity
jump (subshock) indicated in Fig.~\ref{sketch}.

$\bullet$ Strong collisionless plasma shocks with an efficient Fermi
acceleration of energetic particles could generate strong MHD waves
in the upstream and downstream regions and strongly  amplify the
upstream magnetic fields. A distinctive feature of the shock is a
predicted possibility of gas pre-heating in the far upstream region
due to MHD wave dissipation, that can produce an extended filament
of temperature $\gtrsim$ 0.1 keV.

$\bullet$ Shock waves both from the cosmic web formation processes
and those due to cluster merging activity can play an important role
in clusters of galaxies. Direct evidences for such shocks, as traced
by radio relics and the temperature jumps in X-ray observations havebeen found only in a small number of clusters, and thus we need more
observations.

\begin{acknowledgements}
The authors thank ISSI (Bern) for support of the team
``Non-virialized X-ray components in clusters
    of galaxies''. A.M.B. thanks M.Yu. Gustov for his help with hybrid shock
    simulations. He acknowledges the RBRF grant 06-02-16844, 
    a support from RAS Presidium Programmes.  A support from NASA ATP (NNX07AG79G) is acknowledged.
\end{acknowledgements}

\bibliographystyle{aa}
\bibliography{07_bykov}

\end{document}